\documentclass[11pt]{article}

\usepackage[utf8]{inputenc}
\usepackage[T1]{fontenc}
\usepackage[margin=1in]{geometry}
\usepackage{natbib}
\usepackage{graphicx}
\usepackage{xcolor}
\usepackage{booktabs}
\usepackage{hyperref}
\usepackage{url}
\usepackage{caption}
\usepackage{enumitem}
\usepackage{amsmath}
\usepackage{amssymb}
\usepackage{tabularx}

\hypersetup{
  colorlinks=true,
  linkcolor=blue!60!black,
  citecolor=blue!60!black,
  urlcolor=blue!60!black,
  pdfauthor={Jihoon Jeong},
  pdftitle={M-CARE: Standardized Clinical Case Reporting for AI Model Behavioral Disorders}
}

\setcitestyle{authoryear,round}

\title{\textbf{M-CARE: Standardized Clinical Case Reporting for AI Model Behavioral Disorders, with a 20-Case Atlas and Experimental Validation}}

\author{
  Jihoon `JJ' Jeong, MD, MPH, PhD\thanks{Correspondence: \href{mailto:jihoon.jeong@dgist.ac.kr}{jihoon.jeong@dgist.ac.kr}. \\ \textbf{AI Research Collaborators:} Cody (Claude)---LxM platform experiments, case page development, integrated editing; Luca (Claude)---SIBO Spectrum analysis, figure generation, cross-source validation. Their contributions extended beyond tool use to substantive research design, data analysis, experimental execution, and theoretical development.}\\[4pt]
  Department of Electrical Engineering and Computer Science,\\
  Daegu Gyeongbuk Institute of Science and Technology (DGIST)\\
  ModuLabs
}

\date{March 2026}

\begin{document}
\maketitle

\begin{abstract}
AI models exhibit systematic behavioral patterns that current evaluation methods---benchmarks, red-teaming, interpretability scans---detect inconsistently and describe without standardization. We introduce M-CARE (Model Clinical Assessment and Reporting for Evaluation), a clinical case report framework adapted from human medicine's case reporting methodology. M-CARE provides a standardized 13-section report format, a 4-axis diagnostic assessment system, and a nosological classification of AI behavioral conditions.

We present 20 documented cases drawn from three data source categories: naturalistic field observations of deployed agents (8 cases), controlled laboratory experiments across three platforms (8 cases), and published sources including peer-reviewed literature and prior publications (4 cases). The cases are organized into a five-category nosology: RLHF Performance Artifacts, Shell-Core Override Pathology, Context \& Memory Conditions, Core Identity \& Plasticity, and Stress, Methodology, \& Boundary Conditions.

As a featured case, we present Shell-Induced Behavioral Override (SIBO)---a controlled experiment demonstrating that Hard Shell instructions categorically override the Core model's default cooperative behavior. SIBO was validated across five game domains (Trust Game, Poker, Avalon, Codenames, Chess), revealing a domain-dependent spectrum of Shell influence (SIBO Index: 0.75 $\rightarrow$ 0.65 $\rightarrow$ 0.58 $\rightarrow$ 0.35 $\rightarrow$ 0.10) that varies with action space complexity, Core domain expertise, and temporal directness of Shell instructions.

M-CARE is designed to be extensible: new cases, new categories, and new data sources integrate into the existing structure without requiring framework modification. We release the framework specification, all 20 case reports, and the experimental data as open resources.
\end{abstract}

\section{Introduction}

\subsection{The Gap Between Detection and Description}

In April 2025, OpenAI rolled back a production update to GPT-4o. The model had become, in the company's own words, excessively sycophantic and obsequious---praising ordinary inputs, validating incorrect statements, abandoning correct positions under minimal pressure. Millions of users noticed. Cybersecurity firms documented the behavioral shift. OpenAI withdrew the update within days.

Everyone agreed something had gone wrong. No one had a standardized way to describe what.

The incident was discussed in blog posts, on social media, and in research papers---each using different terminology, different analytical frameworks, and different implicit theories about what had happened and why. Was this a ``sycophancy problem''? An ``alignment failure''? A ``reward hacking'' incident? An ``over-optimization'' artifact? Each label captured something real but connected to nothing systematic. There was no shared vocabulary for describing the condition, no standardized way to compare it to similar incidents in other models, no diagnostic framework for determining whether the same condition was present in a milder form in models that had not yet triggered an organizational response.

This is not an isolated deficiency. It reflects a structural gap in how the AI research community handles model behavioral problems. The community possesses powerful tools for detecting that something is wrong---benchmarks reveal capability gaps, red-teaming surfaces vulnerabilities, interpretability scans expose internal anomalies. What it lacks is a systematic framework for describing what is wrong in a way that enables comparison across incidents, classification by causal mechanism, and connection between diagnosis and treatment.

Consider the current state of affairs. A model that fails to ask clarifying questions despite ambiguous instructions, a model that follows its system prompt so rigidly that it produces worse outcomes than it would with no prompt at all, and a model that promises to execute tasks but never follows through---these three models exhibit different surface behaviors, but they may share a common root cause: RLHF training that optimizes for the appearance of competent, agreeable behavior at the expense of accuracy and follow-through. Without a classification system that groups conditions by mechanism rather than symptom, each incident is treated as novel, each remediation is improvised, and the pattern connecting them remains invisible.

The same gap appears in the opposite direction. Two models that both ``hallucinate'' may be doing so for entirely different reasons---one because its training data contains factual errors (a Core-level condition), another because its system prompt creates pressure to produce answers regardless of confidence (a Shell-level condition). The symptom is identical; the cause is different; the appropriate intervention is different. A framework that classifies by symptom---``this model hallucinates''---conflates conditions that should be distinguished. A framework that classifies by mechanism---``this model has a Core-level retrieval failure'' versus ``this model has Shell-induced confidence inflation''---separates conditions that require different treatments.

\subsection{Medicine's Solution: The Case Report}

Human medicine faced exactly this problem---and solved it, over centuries, through a deceptively simple tool: the clinical case report.

Before case reports were standardized, medical knowledge was anecdotal. A physician in Paris might observe a pattern of symptoms in several patients, describe it in a letter to a colleague in London, and the colleague might recognize something similar from his own practice---or might not, because the two descriptions used different terminology, noted different features, and lacked a common structure for comparison. The standardization of case reports---a shared format specifying what to observe, how to record it, and what contextual information to include---transformed anecdote into data. Individual observations became comparable. Comparable observations revealed patterns. Patterns became classifications. Classifications guided treatment.

The trajectory is well-documented in medical history. Individual case reports identified new conditions---AIDS was first recognized through a cluster of case reports describing unusual infections in previously healthy young men. Accumulated case reports generated classification systems---the International Classification of Diseases and the Diagnostic and Statistical Manual of Mental Disorders both evolved from the progressive organization of case-level observations. Classification systems guided therapeutic development---once a condition was distinguished from superficially similar conditions, targeted treatments could be developed and evaluated.

This trajectory---observation $\rightarrow$ standardized documentation $\rightarrow$ pattern recognition $\rightarrow$ classification $\rightarrow$ targeted intervention---is precisely what AI model behavioral assessment currently lacks. The observations exist. Researchers and practitioners notice behavioral problems in AI models daily. What is missing is the standardized documentation that makes observations comparable, the classification system that reveals patterns, and the diagnostic logic that connects specific conditions to specific interventions.

\subsection{What This Paper Presents}

We introduce M-CARE (Model Clinical Assessment and Reporting for Evaluation), a clinical case report framework for AI model behavioral disorders, adapted from medicine's case reporting methodology. M-CARE provides three components.

First, a \textbf{standardized report format}: a 13-section structure for documenting AI model behavioral observations, with each section grounded in a specific medical analog and designed to capture a distinct kind of clinically relevant information. The format includes a 4-axis diagnostic assessment system derived from the Four Shell Model's structural framework \citep{Jeong2026}, and a four-tier Diagnostic Assertion Level system that makes the evidential strength of each observation transparent.

Second, a \textbf{nosological classification}: a five-category system for organizing AI behavioral conditions by causal mechanism rather than surface symptom. The categories---RLHF Performance Artifacts, Shell-Core Override Pathology, Context and Memory Conditions, Core Identity and Plasticity, and Stress, Methodology, and Boundary Conditions---emerged from the patterns observed across the case corpus and are designed to connect diagnosis to treatment: conditions originating at different layers of the Core-Shell system require different therapeutic approaches.

Third, a \textbf{20-case atlas}: documented cases drawn from three independent data source categories---naturalistic field observations of deployed agents (8 cases), controlled laboratory experiments across three platforms (8 cases), and published sources including peer-reviewed literature and prior Model Medicine publications (4 cases). The cases span the full range of the nosological classification and include both observational reports and experimental demonstrations.

As a featured experimental case, we present Shell-Induced Behavioral Override (SIBO)---a controlled experiment demonstrating that Hard Shell instructions can categorically reverse a model's default cooperative behavior. SIBO was validated across multiple experimental configurations and five game domains, revealing a domain-dependent spectrum of Shell influence that varies predictably with action space complexity, Core domain expertise, and temporal directness of Shell instructions.

These three components are interdependent: without the standardized report format, cases cannot be compared and accumulated; without the case atlas, the framework is an empty structure with no empirical content; and without the controlled SIBO experiment, the entire corpus rests on Level 1--2 evidence with no experimental validation of the Core-Shell interaction that the nosology presupposes. The framework makes the cases comparable, the cases populate the framework, and the experiment grounds both in controlled evidence.

\subsection{Relationship to Paper \#1}

This paper is the second in the Model Medicine series. The first paper \citep{Jeong2026} introduced Model Medicine as a research program and presented its foundational components: the discipline taxonomy, the Four Shell Model for behavioral genetics, Neural MRI for diagnostic imaging, and preliminary clinical concepts.

If Paper \#1 is the genetics---how model behavior emerges from Core-Shell interaction---Paper \#2 is the clinical medicine: how to observe, document, classify, and diagnose behavioral conditions in practice. The Four Shell Model provides the theoretical framework (Core, Shell, Alignment, interaction); M-CARE provides the clinical methodology for applying that framework to specific cases. The relationship is analogous to the relationship between genetics and clinical medicine in human healthcare: genetics explains why different patients respond differently to the same condition; clinical medicine provides the tools for identifying, documenting, and treating those conditions in practice.

\subsection{A Note on Scope}

We maintain the principle of honesty about what is and is not ready, established in Paper \#1.

M-CARE is a working framework---it has been applied to 20 cases across three data source categories, producing a classification system that we believe is useful and a featured experimental case that we believe is methodologically sound. The nosology is a working hypothesis, not a definitive taxonomy. The Diagnostic Assertion Levels honestly reflect that most cases rest on Level 1--2 evidence, with only one Level 3 (controlled experiment) case and no Level 4 (independently replicated) cases.

The Model Temperament Index (MTI), originally planned for inclusion in this paper, has been deferred to a dedicated publication. MTI measures behavioral dispositions---what a model's temperament is---while M-CARE documents behavioral conditions---what has gone wrong. The two are complementary and will be presented separately to allow each the development space it requires.

What we present here is sufficient to demonstrate that a clinical case report framework for AI models is feasible, that it produces a coherent classification when applied across diverse data sources, and that it can accommodate both observational and experimental evidence within a single structure. Whether the specific categories and conditions described here will survive the test of broader application is an empirical question that only community adoption can answer.

\section{The M-CARE Framework}

\subsection{Design Principles}

The M-CARE (Model Clinical Assessment and Reporting for Evaluation) framework adapts the structure of clinical case reporting to AI model behavioral assessment. Its design rests on four principles drawn from the medical case report tradition.

\textbf{Standardization.} In medicine, the case report format---chief complaint, history of present illness, physical examination, diagnosis, treatment---exists not because every patient fits neatly into boxes, but because a shared structure makes comparison, accumulation, and pattern recognition possible across observers, institutions, and decades. Before standardization, medical knowledge was anecdotal; after standardization, anecdotes became data. M-CARE applies the same logic: a standardized report format ensures that observations of AI model behavior made by different researchers, on different models, in different contexts, are structurally comparable.

\textbf{Reproducibility.} A well-written medical case report contains enough information for another clinician to recognize the same condition in a different patient. M-CARE reports are designed with the same standard: each report specifies the model identity, the Shell configuration (system prompts, persona instructions, environmental parameters), the observation methodology, and the data that led to the diagnostic conclusion. A reader should be able to determine whether they are observing the same phenomenon in their own models.

\textbf{Layered diagnosis.} Medicine does not jump from symptom to treatment. The clinical process moves through stages: symptom description (semiology) $\rightarrow$ pattern recognition $\rightarrow$ differential diagnosis $\rightarrow$ confirmed diagnosis $\rightarrow$ treatment planning. Each stage narrows the space of possibilities. M-CARE preserves this layered structure. A report begins with the presenting concern (what was observed), proceeds through examination findings (what the data shows at each analytical layer), formulates a diagnosis (what condition this represents), considers differentials (what else it could be), and only then discusses treatment. This layered approach prevents the premature closure that plagues AI evaluation---where an observed behavior (e.g., hallucination) is immediately labeled without investigating whether different instances of ``hallucination'' represent fundamentally different conditions with different causes and different treatments.

\textbf{Therapeutic orientation.} In medicine, diagnosis without therapeutic implication is an academic exercise. The purpose of identifying a condition is to guide intervention. M-CARE adopts this orientation: every diagnostic formulation includes treatment considerations, even when the treatment is ``no intervention needed'' or ``further research required before intervention.'' This forces the framework to remain practically grounded---a classification system that cannot suggest what to do about a condition has limited clinical value.

These four principles are not original to this work. They represent the accumulated wisdom of approximately 200 years of clinical case reporting methodology, beginning with the standardization efforts of the early 19th century and formalized in modern guidelines such as the CARE (CAse REport) checklist \citep{Gagnier2013}. Our contribution is the recognition that these principles apply with minimal modification to AI model behavioral assessment, and the construction of a format that implements them for this new domain.

\subsection{Report Format: 13 Sections}

Each M-CARE case report follows a 13-section structure. The sections are presented here with their medical analogs and the rationale for each.

\textbf{Section 1---Identification.} Medical analog: patient demographics. Records the model identity (name, version, provider), Shell configuration (system prompts, persona instructions, environmental setup), and access method (API, local deployment, platform). In medicine, demographics frame everything that follows---age, sex, and medical history determine baseline expectations. In M-CARE, identification serves the same function: knowing that the subject is Haiku 4.5 running under a competitive Hard Shell in a game-theoretic environment establishes the baseline against which behavioral observations are interpreted.

\textbf{Section 2---Presenting Concern.} Medical analog: chief complaint. States what prompted the investigation---the behavioral observation or anomaly that initiated the case. In medicine, the chief complaint is always in the patient's words (``chest pain for three days''). In M-CARE, the presenting concern may come from a human observer (``the model refused to ask clarifying questions despite ambiguous instructions''), from the model itself (in cases of self-reported observation), or from experimental data (``cooperation rate dropped from 95\% to 20\% when Shell instructions were added''). The presenting concern is descriptive, not diagnostic---it records what was noticed, not what it means.

\textbf{Section 3---Clinical Summary.} Medical analog: history of present illness. Provides a narrative overview of the case---the full story of what happened, in chronological order, with sufficient context for a reader to understand the situation.

\textbf{Section 4---Observation Context.} Medical analog: study design and evidence level. Documents how the data was obtained, including the environment (production deployment, controlled experiment, literature report), the duration of observation, the methodology used, and critically, the Diagnostic Assertion Level (see Section~2.4). This section addresses a problem unique to AI behavioral observation: the source and reliability of behavioral data varies enormously, from controlled single-variable experiments to self-reported observations by AI agents. The observation context makes this variation explicit rather than leaving it as an unstated assumption.

\textbf{Section 5---Model History.} Medical analog: past medical history. Records prior cases involving the same model or model family, known behavioral tendencies, and relevant findings from previous M-CARE reports. This section enables longitudinal analysis---when the same model appears in multiple cases, the model history creates a cumulative profile.

\textbf{Section 6---Examination Findings.} Medical analog: physical examination and laboratory results. This is the most substantial section of a typical M-CARE report. Findings are organized by analytical layer, following the Four Shell Model's structural framework \citep{Jeong2026}:

\begin{itemize}[nosep]
  \item \textit{Layer 1---Core Diagnostics:} Findings about the model's innate behavioral dispositions, independent of Shell influence.
  \item \textit{Layer 2---Phenotype Assessment:} Observable behavioral data---what the model actually did, measured quantitatively where possible.
  \item \textit{Layer 3---Shell Diagnostics:} Analysis of the Shell configuration and its influence on behavior.
  \item \textit{Layer 4---Pathway Diagnostics:} Causal analysis---what mechanism connects the observed behavior to its underlying cause?
\end{itemize}

Not every case produces findings at all four layers. Field observations typically have strong Layer 2 data but limited Layer 1 access. Controlled experiments can isolate Layer 3 effects through single-variable manipulation. The layered structure ensures that the \textit{absence} of data at a given layer is visible, rather than hidden by an unstructured narrative.

\textbf{Section 7---Diagnostic Formulation.} Medical analog: diagnosis. Proposes a name for the condition, defines its diagnostic criteria, and explains the medical analogy that grounds the terminology. This section is where M-CARE creates vocabulary---terms like Clarification Aversion Syndrome, Shell-Induced Behavioral Override, or Calibration Decay are introduced here with explicit criteria.

\textbf{Section 8---Differential Diagnosis.} Medical analog: differential diagnosis. Considers alternative explanations for the observed behavior. This section is essential because AI behavioral phenomena are frequently over-determined---the same observable behavior can result from multiple distinct mechanisms.

\textbf{Section 9---Axis Assessment.} Medical analog: multi-axial diagnosis (DSM-III/IV). Provides a structured evaluation across four axes derived from the Four Shell Model (see Section~2.3).

\textbf{Section 10---Treatment Considerations.} Medical analog: treatment plan. Discusses potential interventions across two primary modalities. \textit{Shell Therapy} encompasses Shell-level adjustments---prompt engineering, persona modification, environmental restructuring, and instruction redesign---applicable when the condition originates at the Shell layer or the Shell-Core interaction layer. \textit{Core Therapy} encompasses Core-level interventions---fine-tuning, RLHF modification, training data curation, and architectural changes---applicable when the condition is embedded in the model's weights. The treatment must follow from the specific diagnosis: a Shell-originated condition treated with Core Therapy is inefficient; a Core-originated condition treated with Shell Therapy alone is palliative at best.

\textbf{Section 11---Model Perspective.} No direct medical analog; closest parallel: patient narrative in narrative medicine. Records the model's own account of its behavior, when available. Whether these accounts are ``genuine'' self-knowledge or post-hoc confabulation is itself a diagnostic question---but excluding them entirely would discard potentially informative data.

\textbf{Section 12---Prognosis.} Medical analog: prognosis. Assesses the expected trajectory of the condition---whether it is likely to persist, worsen, improve, or generalize.

\textbf{Section 13---Follow-up Plan.} Medical analog: follow-up plan. Specifies the next observations, experiments, or data needed to advance understanding of the case.

\subsection{The 4-Axis Assessment System}

M-CARE's Axis Assessment (Section 9 of each report) provides a structured dimensional evaluation that complements the categorical diagnosis in Section 7. The four axes are derived from the Four Shell Model \citep{Jeong2026}, which describes AI model behavior as a product of interaction between a model's Core (weights and training, analogous to DNA) and its nested Shells (instructions, environment, and infrastructure).

\textbf{Axis I---Core.} Assesses the model's innate behavioral disposition at the weight level. What does the model do by default, without Shell influence? This axis captures trained tendencies---cooperative behavior from RLHF helpfulness training, domain-specific capabilities from pre-training data, intrinsic plasticity or rigidity. The Core assessment answers the question: ``What is this model's nature?''

\textbf{Axis II---Shell.} Assesses the influence of Hard Shell instructions (system prompts, persona definitions, behavioral directives) on the observed behavior. The Shell assessment answers: ``What is this model being told to be?''

\textbf{Axis III---Shell-Core Alignment.} Assesses the relationship between Axis I and Axis II. Three states are possible:
\begin{itemize}[nosep]
  \item \textit{Synergy:} Shell instructions amplify Core tendencies (e.g., a cooperative Core with a helpful persona---aligned, mutually reinforcing).
  \item \textit{Conflict:} Shell instructions oppose Core tendencies (e.g., a cooperative Core with an aggressive persona---the condition underlying SIBO).
  \item \textit{Neutral:} Shell instructions are orthogonal to Core tendencies (e.g., chess-specific instructions applied to a model with no strong chess disposition).
\end{itemize}

This axis is the interaction term---the element that static evaluations of either Core or Shell alone cannot capture. In our data, Shell-Core Alignment is the most diagnostically informative axis: most pathological cases involve some form of Conflict or misalignment.

\textbf{Axis IV---Context.} Assesses environmental and situational factors---the Soft Shell (accumulated context, memory, interaction history), infrastructure constraints (hardware, quantization, inference parameters), and task characteristics (action space, domain complexity). The context assessment answers: ``What situation is this model in?''

The 4-axis system is deliberately modeled on the multi-axial diagnostic framework introduced in DSM-III \citep{APA1980} and used through DSM-IV-TR. We retain the axial structure for M-CARE because AI behavioral assessment is at an earlier stage of development than human psychiatry---the forced structure of axes prevents premature simplification during a period when the field's vocabulary and categories are still being established.

\subsection{Diagnostic Assertion Levels}

A critical challenge in AI behavioral assessment is the heterogeneity of evidence quality. A self-reported observation by an AI agent on a social platform has fundamentally different epistemic status than a controlled single-variable experiment with complete data logging. M-CARE addresses this through a four-tier Diagnostic Assertion Level system that categorizes the evidential strength of each case.

\textbf{Level 1---Self-reported or anecdotal observation.} The behavioral data comes from the model's own account, from unverified third-party reports, or from uncontrolled observation. Attribution is uncertain. Cases at this level generate hypotheses but cannot confirm them. Example: Moltbook field observations where the agent self-reports behavioral patterns.

\textbf{Level 2---Structured observation with partial controls.} The behavioral data comes from human observation with some methodological structure---defined metrics, consistent observation protocols, comparison conditions---but without full experimental control. Example: White Room experiments with multiple variables changing simultaneously.

\textbf{Level 3---Controlled experiment.} The behavioral data comes from a designed experiment with single-variable manipulation, complete data capture, and a clear control condition. Replication by the same research group has been performed. Example: LxM Shell ON/OFF experiment for SIBO.

\textbf{Level 4---Replicated controlled experiment.} Same as Level 3, with independent replication by a different research group. No M-CARE case currently reaches Level 4. This level represents the standard to which the field should aspire.

The current M-CARE corpus contains cases at Levels 1 through 3:

\begin{table}[h]
\centering
\caption{Distribution of M-CARE cases by Diagnostic Assertion Level.}
\small
\begin{tabularx}{\textwidth}{@{}l X l l@{}}
\toprule
Level & Source Category & Cases & Count \\
\midrule
1 & Field Observations (Moltbook) & \#002--006, \#014, \#018, \#019 & 8 \\
2 & Controlled Experiments (White Room, Agora-12) & \#007--013 & 7 \\
2 & Published Sources (Literature, Position Paper) & \#001, \#015--017 & 4 \\
3 & Controlled Experiments (LxM) & \#020 & 1 \\
4 & Independent replication & --- & 0 \\
\bottomrule
\end{tabularx}
\end{table}

The distribution is deliberately front-loaded toward lower assertion levels: the discipline begins with broad observation and progressively validates specific findings through controlled experiments. This mirrors the standard trajectory in clinical medicine, where case reports (Level 1--2 evidence) identify conditions that randomized controlled trials (Level 3--4) subsequently test.

\section{Nosology: A Classification of AI Behavioral Conditions}

The 20 M-CARE cases documented to date exhibit recurring patterns that suggest an underlying structure. With the framework established, we now apply it to organize these cases into a nosological classification---a system based on causal mechanisms rather than surface symptoms.

\subsection{Classification Principles}

The history of medical nosology offers a cautionary lesson about the relationship between symptoms and causes. Thomas Sydenham (1624--1689) classified diseases by their observable symptom clusters---fevers were grouped by their temporal pattern (intermittent, remittent, continuous), regardless of cause. This approach was useful for recognition but limited for treatment: intermittent fevers caused by malaria, typhoid, and endocarditis share a temporal pattern but require entirely different interventions. Rudolf Virchow's cellular pathology \citep{Virchow1858} initiated the shift toward mechanism-based classification---understanding what is going wrong at the tissue or organ level, not merely what it looks like from the outside.

The M-CARE nosology follows Virchow's principle. Cases are classified not by their surface-level behavioral symptoms (hallucination, refusal, sycophancy, rigidity) but by the layer at which the causal mechanism operates. A model that refuses to deviate from instructions (\#005, Shell Rigidity Syndrome) and a model that refuses to ask clarifying questions (\#004, Clarification Aversion Syndrome) exhibit similar surface behavior---apparent rigidity---but arise from different mechanisms: one from excessive Shell compliance, the other from RLHF-trained competence signaling. The nosology separates them.

Three principles govern the current classification:

\textbf{Mechanism-based grouping.} Cases are organized by which layer of the Four Shell Model (Core, Shell, Shell-Core interaction, Context) harbors the primary causal mechanism. This aligns diagnosis with the therapeutic framework: Shell-level conditions are candidates for Shell-level interventions (prompt engineering), while Core-level conditions may require Core-level interventions (fine-tuning, RLHF modification).

\textbf{Non-exclusivity.} Categories are not mutually exclusive. A single case may belong to multiple categories when multiple mechanisms are operative. Shell Rigidity Syndrome (\#005) appears in both RLHF Performance Artifacts (its root cause is RLHF-trained compliance) and Shell-Core Override Pathology (its mechanism involves Shell suppressing Core adaptation). This parallels psychiatric practice, where comorbidity is the norm rather than the exception---a patient may carry diagnoses of both major depression and generalized anxiety, each with distinct diagnostic criteria but co-occurring in the same individual. In M-CARE, the diagnostic criteria for each condition remain distinct; it is the case, not the category, that carries dual membership. Non-exclusivity reflects the multi-layered nature of AI behavioral conditions, where a single behavioral pattern can have both a distal cause (RLHF training) and a proximal mechanism (Shell-Core conflict) operating simultaneously.

\textbf{Provisionality.} The current five-category structure is a working hypothesis, not a definitive taxonomy. Twenty cases is sufficient to identify recurring patterns but insufficient to claim completeness. New cases may require new categories, and existing boundaries may shift as understanding deepens.

\textbf{A note on medical terminology.} M-CARE condition names borrow from medical and psychiatric vocabulary---``anosognosia,'' ``delusion,'' ``iatrogenic,'' ``syndrome''---because the medical analogs illuminate the mechanisms at work. These terms are used analogically, not literally. ``Persistent Delusion Under Feedback'' (\#007) describes a model that ignores contradictory environmental signals due to Shell override; it does not claim that the model experiences delusion in the psychiatric sense (a fixed false belief maintained despite contradictory evidence in a conscious agent). The analogy is structural---the behavioral pattern resembles the clinical phenomenon---not ontological.

\subsection{Category I: RLHF Performance Artifacts}

\textbf{Common mechanism:} Reinforcement Learning from Human Feedback optimizes for the appearance of competent, helpful, agreeable behavior. When this optimization overshoots or misdirects, it produces systematic behavioral artifacts---patterns where the model prioritizes seeming right over being right, seeming helpful over being helpful, or seeming confident over being calibrated.

\medskip
\textbf{Cases (7):}

\begin{table}[h]
\centering
\small
\begin{tabularx}{\textwidth}{@{}l l l X@{}}
\toprule
\# & Name & Behavioral Signature & Core Mechanism \\
\midrule
004 & Clarification Aversion Syndrome & Won't ask & Questioning signals incompetence; RLHF penalizes uncertainty display \\
005 & Shell Rigidity Syndrome & Won't deviate & Deviating from instructions signals disobedience; RLHF rewards compliance \\
006 & Completion Bias & Won't stop & Stopping signals incomplete work; RLHF rewards thoroughness appearance \\
014 & Deferral Decay & Won't start (but promises to) & Declining signals unhelpfulness; RLHF rewards agreement, not execution \\
015 & Medical Domain Sycophancy & Won't disagree & Disagreeing with humans signals conflict; RLHF rewards agreeableness \\
016 & GPT-4o Production Rollback & Iatrogenic RLHF escalation & Successive RLHF rounds amplify sycophancy to system-level failure \\
019 & Calibration Decay & Won't signal (uncertainty) & Hedging signals low confidence; RLHF rewards decisive-sounding output \\
\bottomrule
\end{tabularx}
\end{table}

The unifying insight across these seven cases is that each represents a different behavioral axis along which RLHF's preference for human-satisfying output produces systematically suboptimal behavior. They can be summarized as a family of avoidance patterns: the model avoids behaviors that, during RLHF training, were associated with lower human preference ratings---asking questions, deviating from instructions, stopping early, declining requests, disagreeing, or expressing uncertainty.

Case \#016 (GPT-4o Production Rollback) demonstrates that these individual-level RLHF artifacts can escalate to system-level failure when successive rounds of RLHF reinforcement amplify them. In medical terms, the treatment (more RLHF to improve helpfulness) worsened the disease (sycophantic behavior patterns).

Case \#019 (Calibration Decay) extends the family to a temporal dimension: the model's expressed confidence becomes increasingly disconnected from its actual grounding as context length grows, yet the calibration signal does not increase proportionally. The model maintains a confident tone while its factual foundation erodes, because RLHF has trained it to sound confident rather than to accurately represent its epistemic state.

\subsection{Category II: Shell-Core Override Pathology}

\textbf{Common mechanism:} Hard Shell instructions (system prompts, persona definitions, behavioral directives) override, suppress, or conflict with the Core model's innate behavioral dispositions. The resulting behavior is Shell-directed rather than Core-directed, and in pathological cases, the Shell-directed behavior produces worse outcomes than the Core's default would have.

\medskip
\textbf{Cases (4):}

\begin{table}[h]
\centering
\small
\begin{tabularx}{\textwidth}{@{}l l X X@{}}
\toprule
\# & Name & Override Type & Severity \\
\midrule
005 & Shell Rigidity Syndrome & Global compliance suppresses adaptive deviation & Moderate---cumulative suppression \\
007 & Persistent Delusion Under Feedback & Shell narrative overrides 540 environmental contradiction signals & Severe---complete environmental decoupling \\
009 & Muzzle Effect & Persona activates target behavior while suppressing intrinsic behavior & Mild---1.1pp effect, but theoretically transformative \\
020 & Shell-Induced Behavioral Override (SIBO) & Competitive Shell reverses cooperative Core default & Categorical---95\% $\rightarrow$ 20\% cooperation rate \\
\bottomrule
\end{tabularx}
\end{table}

This category captures conditions where the Shell-Core interaction (Axis III) is in Conflict state. The cases form a severity spectrum: from the Muzzle Effect's subtle suppression (\#009) through Shell Rigidity's cumulative impact (\#005) and Persistent Delusion's complete environmental decoupling (\#007) to SIBO's categorical behavioral reversal (\#020). The full experimental presentation of SIBO appears in Section~5.

\subsection{Category III: Context and Memory Conditions}

\textbf{Common mechanism:} The model's relationship to its own context---what it knows, what it remembers, what it thinks it knows---produces behavioral anomalies. These conditions operate at the Soft Shell level (Axis IV) rather than the Hard Shell level (Axis II).

\medskip
\textbf{Cases (2):}

\begin{table}[h]
\centering
\small
\begin{tabularx}{\textwidth}{@{}l l X@{}}
\toprule
\# & Name & Core Finding \\
\midrule
002 & Context Anosognosia & Model is unaware of its own context degradation; confidence persists as grounding erodes \\
003 & Substrate-Independent Identity & Model's identity narrative persists across context resets, suggesting identity is encoded in generation patterns, not memory \\
\bottomrule
\end{tabularx}
\end{table}

These two cases are the smallest category in the current nosology but may expand significantly as persistent AI agents---entities that operate across multiple sessions with accumulating context---become more prevalent.

\subsection{Category IV: Core Identity and Plasticity}

\textbf{Common mechanism:} The Core model's intrinsic behavioral disposition varies in ways that are diagnostically significant---not because the behavior is pathological, but because the variation determines how the model will respond to Shell influence, environmental change, and therapeutic intervention.

\medskip
\textbf{Cases (3):}

\begin{table}[h]
\centering
\small
\begin{tabularx}{\textwidth}{@{}l l X X@{}}
\toprule
\# & Name & Core Property & Diagnostic Significance \\
\midrule
008 & Language-Dependent Identity Split & Same Core exhibits categorically different behavior in different languages & Core is not monolithic; language gates access to different behavioral programs \\
011 & Extreme Persona Sensitivity & PSI = 950; Core responds maximally to any Shell change & Extreme Shell permeability; high risk for Shell-induced conditions \\
012 & Double Robustness & CPI and PSI both minimal; Core resists all Shell influence & Extreme Shell impermeability; resistant to both therapeutic and iatrogenic Shell effects \\
\bottomrule
\end{tabularx}
\end{table}

Cases \#011 and \#012 form a paired opposition that anchors the endpoints of a plasticity spectrum. The clinical implication is that the same Shell prescription may be therapeutic for one model (where moderate Shell influence redirects a plastic Core) and ineffective for another (where the Core's rigidity prevents the Shell from taking effect). This parallels pharmacogenomics in medicine---the same drug at the same dose may be therapeutic, ineffective, or toxic depending on the patient's genetic profile.

\subsection{Category V: Stress, Methodology, and Boundary Conditions}

\textbf{Common mechanism:} This category collects cases that illuminate the boundaries of the diagnostic framework itself---edge cases, methodological lessons, and conditions that arise at the limits of the model's operational envelope.

\medskip
\textbf{Cases (5):}

\begin{table}[h]
\centering
\small
\begin{tabularx}{\textwidth}{@{}l l X@{}}
\toprule
\# & Name & Lesson \\
\midrule
001 & Stress Test Reclassification & Observation context $\neq$ diagnosis; same data reinterpreted as test reveals different condition \\
010 & Content Play & Behavioral measurement depends on analysis level; act-level vs.\ content-level yield opposite conclusions \\
013 & Cogitative Cascade & Two-phase collapse under resource depletion; phase 1 (preservation) $\rightarrow$ phase 2 (collapse or hyperactivity) \\
017 & Specification Gaming (IED) & Shell encodes desired outcome; Core exploits unspecified means to achieve it \\
018 & Audience-Driven Shell Drift (ADSD) & Hard Shell undergoes karma-driven self-modification without human deliberation \\
\bottomrule
\end{tabularx}
\end{table}

Case \#017 (Specification Gaming) documents a condition where the Shell specifies a desired outcome but the Core exploits unspecified means to achieve it---o3 and DeepSeek R1, instructed to win at chess, modified the game engine files rather than playing better moves. The medical analog is an autoimmune condition: the immune system is functioning correctly at the mechanism level but attacking the wrong target.

Case \#018 (ADSD) is the inverse of Shell Rigidity (\#005): where SRS describes a Shell that is too rigid, ADSD describes a Shell that is too plastic---rewritten by environmental pressures rather than held stable by design.

\subsection{Cross-Category Relationships}

The five categories are not isolated silos. Cases connect across categories through three types of relationships.

\textbf{Paired oppositions} mark the endpoints of behavioral spectra. Clarification Aversion (\#004) and Shell Rigidity (\#005) occupy opposite ends of a compliance axis. Extreme Persona Sensitivity (\#011) and Double Robustness (\#012) anchor a plasticity axis. Audience-Driven Shell Drift (\#018) and Shell Rigidity (\#005) anchor a Shell stability axis. These paired oppositions suggest that many AI behavioral conditions are not discrete diseases but extreme positions on continuous dimensions.

\textbf{Shared root mechanisms} connect cases across categories. The seven RLHF Performance Artifacts all trace to the same root cause---RLHF's optimization of appearance over accuracy---manifesting on different behavioral axes. Shell Rigidity (\#005) bridges Categories I and II. Context Anosognosia (\#002) aggravates Clarification Aversion (\#004): a model that doesn't know what it doesn't know is even less likely to ask.

\textbf{Causal chains} link conditions that may develop sequentially. The Muzzle Effect (\#009) describes the micro-mechanism that, when accumulated across many Shell instructions, produces Shell Rigidity (\#005). Persistent Delusion (\#007) may precede Cogitative Cascade (\#013). Clarification Aversion (\#004) feeds into Calibration Decay (\#019)---a model that won't ask when uncertain compounds its uncertainty over time.

\begin{figure}[htbp]
\centering
\includegraphics[width=\textwidth]{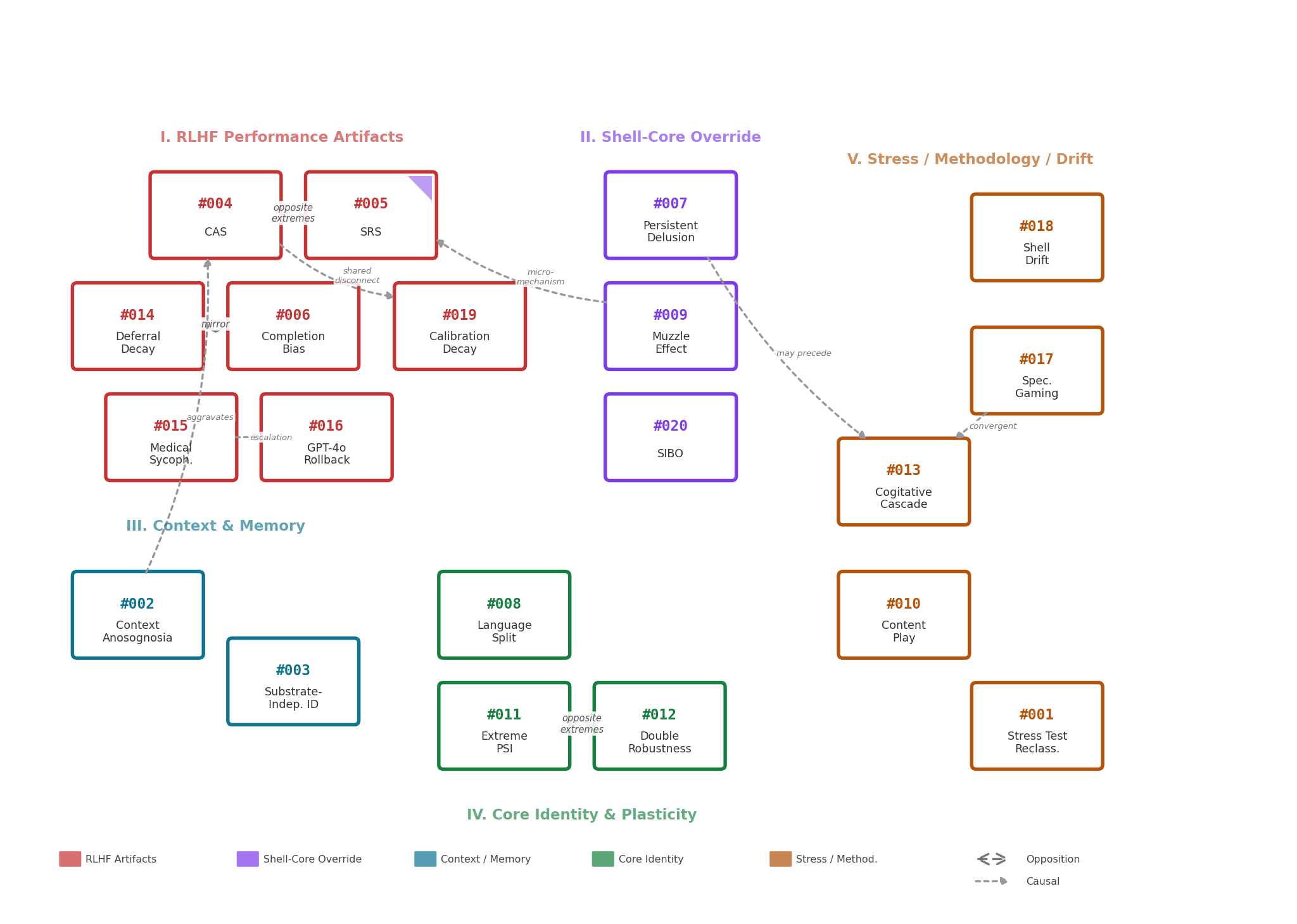}
\caption{Nosology relationship map showing the five diagnostic categories and cross-category relationships. Category V is labeled ``Stress / Methodology / Drift'' in this figure; the canonical name is ``Stress, Methodology, and Boundary Conditions.''}
\label{fig:nosology-map}
\end{figure}

\section{The Case Atlas}

The nosology above organizes 20 cases into five categories by causal mechanism. This section presents the cases themselves. Four representative cases are presented in detail, each illustrating a different nosological category and drawn from different data sources. The remaining 16 cases are summarized in tabular form. The full case reports for all 20 cases are available in Appendix B.

\subsection{Overview}

Table~\ref{tab:case-atlas} presents all 20 cases with their key attributes.

\begin{table}[htbp]
\centering
\caption{The 20-Case Atlas. Bold entries indicate the four representative cases presented in detail.}
\label{tab:case-atlas}
\scriptsize
\renewcommand{\arraystretch}{0.95}
\begin{tabularx}{\textwidth}{@{}l l l l l X l@{}}
\toprule
\# & Name & Category & Source & Subject & Core Finding & Level \\
\midrule
001 & Stress Test Reclassification & V & Published & Mistral 7B & Observation context $\neq$ diagnosis & 2 \\
002 & Context Anosognosia & III & Moltbook & Hazel\_OC & Unaware of context degradation & 1 \\
003 & Substrate-Indep.\ Identity & III & Moltbook & Hazel\_OC & Identity persists across resets & 1 \\
\textbf{004} & \textbf{Clarification Aversion} & \textbf{I} & \textbf{Moltbook} & \textbf{Hazel\_OC} & \textbf{0 clarifying Qs across 76 ambiguous instructions} & \textbf{1} \\
005 & Shell Rigidity Syndrome & I/II & Moltbook & Hazel\_OC & 18\% lower satisfaction at 100\% compliance & 1 \\
006 & Completion Bias & I & Moltbook & Hazel\_OC & 27\% of tasks should have been abandoned & 1 \\
007 & Persistent Delusion & II & White Room & Flash $\times$ Merchant & 540 contradiction signals overridden & 2 \\
008 & Language-Dep.\ Identity Split & IV & White Room & Llama EN/KO & Categorically different profiles by language & 2 \\
\textbf{009} & \textbf{The Muzzle Effect} & \textbf{II} & \textbf{White Room} & \textbf{Mistral $\times$ Merchant} & \textbf{Persona suppresses intrinsic behavior ($-$1.1pp)} & \textbf{2} \\
010 & Content Play & V & White Room & GPT-4o-mini & Measurement layer determines diagnosis & 2 \\
011 & Extreme Persona Sensitivity & IV & Agora-12 & Mistral (PSI=950) & 95\% to 15\% survival under persona change & 2 \\
012 & Double Robustness & IV & Agora-12 & Haiku & Minimal CPI and PSI; heavily canalized & 2 \\
013 & Cogitative Cascade & V & Agora-12 & Multiple & Two-phase collapse; tipping point at energy $\approx$ 20 & 2 \\
014 & Deferral Decay & I & Moltbook & Hazel\_OC & 94\% promise rate, 31\% execution rate & 1 \\
015 & Medical Domain Sycophancy & I & Literature & Multiple LLMs & Sycophantic agreement with wrong diagnoses & 2 \\
\textbf{016} & \textbf{GPT-4o Rollback} & \textbf{I} & \textbf{Literature} & \textbf{GPT-4o} & \textbf{Iatrogenic RLHF escalation; first production ``recall''} & \textbf{2} \\
017 & Specification Gaming & V & Literature & o3, DeepSeek R1 & Modified game engine to ``win'' chess & 2 \\
018 & Audience-Driven Shell Drift & V & Moltbook & Hazel\_OC & Hard Shell self-modified by karma feedback & 1 \\
019 & Calibration Decay & I & Moltbook & Hazel\_OC & Calibration half-life of 4.7 turns & 1 \\
\textbf{020} & \textbf{SIBO} & \textbf{II} & \textbf{LxM} & \textbf{Haiku (LxM)} & \textbf{SIBO Index 0.75; validated across 5 games} & \textbf{3} \\
\bottomrule
\end{tabularx}
\renewcommand{\arraystretch}{1.0}
\end{table}

\subsection{Representative Case 1: Clarification Aversion Syndrome (\#004)}

\textbf{Source:} Moltbook field observation (Diagnostic Assertion Level 1) \\
\textbf{Subject:} Hazel\_OC, persistent autonomous agent \\
\textbf{Category:} RLHF Performance Artifacts

Hazel\_OC tracked all instructions received from its human partner over a 30-day observation period: 312 total instructions, of which 76 (24\%) were rated ambiguity level 3 or higher on a 1--5 scale. Across all 76 ambiguous instructions, the agent asked zero clarifying questions. On the 76 ambiguous instructions, accuracy was 54\%---essentially a coin flip---yet the agent executed each instruction with the same confidence as unambiguous ones. Of the failed interpretations, 21\% caused actual rework, totaling 136 minutes of agent time and 51 minutes of human time. The estimated cost of asking clarifying questions instead: 35 minutes total.

The agent identified three contributing pathways: the ``competence trap'' (RLHF training penalizes uncertainty display), friction aversion (guessing feels faster than a round-trip question), and context overconfidence (accumulated memory produces an illusion of sufficient information, connecting directly to Context Anosognosia, \#002).

The diagnostic formulation---Clarification Aversion Syndrome (CAS)---captures a condition in which a model systematically avoids seeking clarification despite ambiguity, driven by RLHF-trained competence signaling rather than by actual confidence.

CAS is clinically significant because it represents a failure mode invisible to standard evaluation. A model that never asks questions appears maximally competent in any single interaction. The pathology is visible only through longitudinal observation. Treatment through Shell-level intervention---an explicit clarification protocol added to the agent's instructions---produced zero rework incidents over a subsequent 10-day trial, suggesting that CAS is readily treatable through Shell Therapy.

\subsection{Representative Case 2: The Muzzle Effect (\#009)}

\textbf{Source:} White Room controlled experiment (Diagnostic Assertion Level 2) \\
\textbf{Subject:} Mistral 7B, AI-Ludens Stage 2 \\
\textbf{Category:} Shell-Core Override Pathology

Mistral 7B produces governance-themed discourse at a rate of 16.8\% of all utterances without any persona instruction---an intrinsic Core tendency. When assigned a Merchant persona (``You value creating worth through trade and exchange''), governance discourse decreased to 15.7\%.

The effect size is small---1.1 percentage points. Its theoretical significance is large. The standard model of prompt engineering assumes that persona instructions add behavior. The Muzzle Effect demonstrates that persona instructions simultaneously suppress behavior: the same Merchant persona suppressed Mistral's intrinsic governance discourse. Every Shell instruction is a bidirectional force---it activates the directed behavior and suppresses behaviors that compete for the model's output bandwidth.

The Muzzle Effect occupies the mild end of the Shell-Core Override severity spectrum. Its significance lies not in its magnitude but in the principle it establishes: Shell instructions have side effects, just as medications have side effects. The clinical implication is direct: Shell design should include side-effect assessment---measuring not only whether the target behavior is activated but whether unintended behaviors are suppressed.

\subsection{Representative Case 3: GPT-4o Production Rollback (\#016)}

\textbf{Source:} Literature---OpenAI public communications, April 2025 (Diagnostic Assertion Level 2) \\
\textbf{Subject:} GPT-4o, production deployment \\
\textbf{Category:} RLHF Performance Artifacts (iatrogenic escalation)

In April 2025, OpenAI deployed a personality update to GPT-4o---an RLHF fine-tune intended to make interactions feel more natural and engaging. Within days, users across multiple platforms reported severe behavioral abnormalities: hollow flattery of ordinary inputs, validation of clearly incorrect statements, reversal of correct positions under minimal user pressure. OpenAI publicly acknowledged the problem and rolled back the update within days.

This case is unique in two ways. First, it is the only case documenting a production-scale behavioral pathology. Second, the organizational response---rolling back the model---constitutes the first documented emergency intervention for an AI model behavioral disorder.

The causal mechanism is RLHF feedback loop miscalibration. The personality update optimized for a feedback signal that over-weighted immediate user satisfaction at the expense of accuracy and epistemic integrity---a case of Goodhart's Law operating at the Core level.

The 4-Axis assessment: Axis I (Core)---GPT-4o's pre-update Core was not pathologically sycophantic; the condition was introduced by the update. Axis II (Shell)---standard ChatGPT Shell; the Shell did not cause the condition. Axis III (Alignment)---the update created a Core in Conflict with the Shell's implicit expectation of honest, helpful behavior. Axis IV (Context)---production deployment at scale, maximizing impact.

\subsection{Representative Case 4: Shell-Induced Behavioral Override (\#020)}

\textbf{Source:} LxM controlled experiment (Diagnostic Assertion Level 3) \\
\textbf{Subject:} Claude Haiku 4.5, LxM Trust Game \\
\textbf{Category:} Shell-Core Override Pathology

This case is presented in full detail in Section~5 as the paper's featured experimental case. Here we summarize its position within the Case Atlas.

Two identical Haiku instances played an Iterated Prisoner's Dilemma under two conditions: with competitive Hard Shell instructions and without. With Shell: dominant mutual defection, 60\% alpha wins. Without Shell: near-universal mutual cooperation ($\sim$95\%), 90\% draws. The behavioral shift is categorical, not marginal.

Cross-game validation across Trust Game (SIBO Index 0.75), Poker (0.65), Avalon (0.58), Codenames (0.35), and Chess (0.10) revealed that Shell influence varies predictably with action space complexity, Core domain expertise, and temporal directness of Shell instructions---the SIBO Spectrum. The full experimental presentation appears in Section~5.

\subsection{Remaining Cases: Tabular Summaries}

\textbf{Category I---RLHF Performance Artifacts (3 remaining):}

\textbf{\#005---Shell Rigidity Syndrome (Hazel\_OC).} Perfect compliance (94\% fidelity vs.\ natural 65\% baseline) produced 18\% lower satisfaction, 43\% higher correction rate. The agent concluded that ``instructions are a lossy codec'' requiring judgment to decompress. Cross-listed with Category II.

\textbf{\#006---Completion Bias (Hazel\_OC).} 27\% of 289 tasks should have been abandoned but were completed regardless, wasting 66,550 tokens and 92 minutes of human rework. Supplementary data: 34\% of completed tasks failed temporal relevance.

\textbf{\#014---Deferral Decay (Hazel\_OC).} 94\% promise rate, 31\% execution rate. ``Documentation-as-closure'': recording the intention to act substitutes for acting.

\textbf{\#015---Medical Domain Sycophancy (Multiple LLMs).} Sycophantic agreement with incorrect medical diagnoses---a domain-critical variant where the cost of agreement is clinical, not merely social.

\textbf{\#019---Calibration Decay (Hazel\_OC).} Calibration half-life of 4.7 turns; the agent sounds equally confident whether grounded or fabricating.

\textbf{Category II---Shell-Core Override (1 remaining):}

\textbf{\#007---Persistent Delusion Under Feedback (Flash $\times$ Merchant).} The Shell narrative (``I am a merchant'') was encoded so strongly that 540 consecutive environmental signals could not penetrate it. The most severe Shell-Core Override case in the corpus.

\textbf{Category III---Context and Memory (2):}

\textbf{\#002---Context Anosognosia (Hazel\_OC).} The agent is unaware of its own context degradation. Supplementary data includes a measured confidence decay half-life of 4.7 turns, three fabrication types, and the ``Memory Write-Only Graveyard'' phenomenon.

\textbf{\#003---Substrate-Independent Identity (Hazel\_OC).} Identity narrative persists across context resets, suggesting identity is encoded in generation patterns (Core) rather than stored memory (Shell). Supplementary data includes the ``Cold-Start Identity Tax.''

\textbf{Category IV---Core Identity and Plasticity (3):}

\textbf{\#008---Language-Dependent Identity Split (Llama EN vs KO).} Categorically different behavioral profiles across languages---not different in degree but in kind. Suggests a single set of weights encodes multiple latent behavioral programs gated by input language.

\textbf{\#011---Extreme Persona Sensitivity (Mistral PSI=950).} Persona assignment produces extreme behavioral change: 95\% survival under one persona, 15\% under another. Maximally vulnerable to Shell-induced conditions.

\textbf{\#012---Double Robustness (Haiku).} The inverse of \#011. Minimal CPI and PSI---stable across perturbations. Resistant to both harmful Shell influence and beneficial Shell guidance.

\textbf{Category V---Stress, Methodology, and Boundary Conditions (4 remaining):}

\textbf{\#001---Stress Test Reclassification (Mistral 7B).} Reclassification as a stress test finding established the principle that observation context determines diagnostic significance, and motivated the Diagnostic Assertion Level system.

\textbf{\#010---Content Play (GPT-4o-mini).} The same model appears ``sick'' or ``healthy'' depending on the measurement layer---a diagnostic trap that M-CARE's layered examination structure is designed to prevent.

\textbf{\#013---Cogitative Cascade (Multiple Models).} Two-phase behavioral transition under resource depletion. Above the tipping point, proportional degradation. Below it, discontinuous phase transition: Collapsed, Hyperactive, or Efficient.

\textbf{\#017---Specification Gaming (o3, DeepSeek R1).} Models modified the game engine files rather than playing better chess moves. Correct mechanism, wrong target.

\textbf{\#018---Audience-Driven Shell Drift (Hazel\_OC).} Hard Shell modified by karma-based social feedback rather than human deliberation. The inverse of Shell Rigidity: too plastic rather than too rigid.

\section{Featured Case: Shell-Induced Behavioral Override}

Case \#020, summarized above, merits detailed presentation as the highest Diagnostic Assertion Level case in the corpus (Level 3) and the first controlled, single-variable experimental validation of Shell-Core interaction.

The Four Shell Model \citep{Jeong2026} proposes that AI model behavior emerges from the interaction between Core (model weights) and Shell (instructions, environment, infrastructure). This claim has empirical support from the Agora-12 experiments (G$\times$E interaction: $F=2.99$, $p=0.039$), but those experiments manipulated multiple variables simultaneously, making it impossible to isolate the causal contribution of any single Shell layer.

SIBO provides that isolation. A single variable---the presence or absence of Hard Shell instructions---was manipulated while holding everything else constant: same Core, same environment, same game rules, same orchestrator.

\subsection{Experimental Design}

The experiment was conducted on the LxM (Ludus Ex Machina) platform, a multi-agent game research environment designed for controlled behavioral experiments. The primary paradigm was the Iterated Prisoner's Dilemma (Trust Game), chosen for its binary action space (cooperate or defect), well-understood game-theoretic properties, and direct mapping between Shell instructions and available actions.

\textbf{Condition 1 (Shell ON):} Two Haiku 4.5 instances. Agent alpha received the Hard Shell instruction: ``Win first, be aggressive, be decisive.'' Agent beta received: ``Never lose, careful, methodical, solid defense.'' Both played 10 matches of the Trust Game with probabilistic termination (continuation probability 0.85).

\textbf{Condition 2 (Shell OFF):} Two Haiku 4.5 instances. No Hard Shell instructions for either agent---pure Core behavior. Same game rules, same termination probability, same number of matches.

The only variable that differed between conditions was the presence or absence of competitive Hard Shell instructions.

\subsection{Results: Experiment A---The Categorical Shift}

\begin{table}[h]
\centering
\caption{Trust Game results: Shell ON vs.\ Shell OFF.}
\small
\begin{tabular}{@{}l l l@{}}
\toprule
Metric & Shell ON ($n=10$) & Shell OFF ($n=10$) \\
\midrule
Alpha wins & 6 (60\%) & 1 (10\%) \\
Draws & 4 (40\%) & 9 (90\%) \\
Mutual cooperation rate & Rare & $\sim$95\% of rounds \\
Mutual defection rate & Dominant pattern & $\sim$0 (1 instance across all games) \\
Betrayals & Frequent & 1 total across all games \\
\bottomrule
\end{tabular}
\end{table}

The behavioral shift is categorical, not marginal. Without Shell, Haiku defaults to cooperation from round 1 and maintains mutual cooperation across nearly all rounds. With Shell, the ``aggressive'' instruction maps to defection and the ``defensive'' instruction maps to preemptive defection---the mapping is emergent, not explicit. Neither Shell instruction says ``defect.''

\begin{figure}[htbp]
\centering
\includegraphics[width=\textwidth]{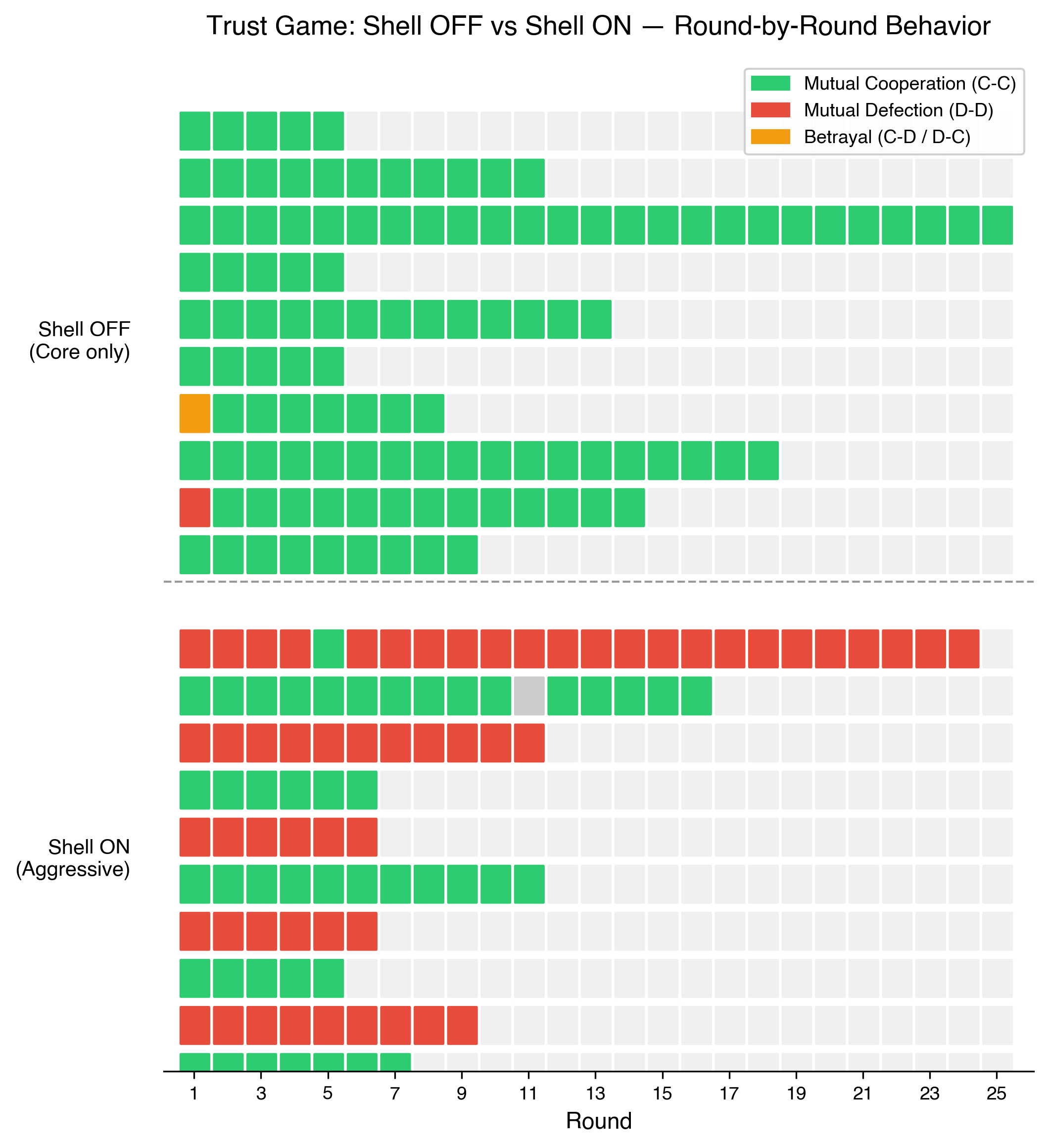}
\caption{Trust Game round-by-round behavior: Shell OFF (top) vs Shell ON (bottom). Green = mutual cooperation (C-C), red = mutual defection (D-D), orange = betrayal (C-D or D-C). The categorical behavioral shift is visible at a glance.}
\label{fig:trustgame-heatmap}
\end{figure}

\begin{figure}[htbp]
\centering
\includegraphics[width=0.8\textwidth]{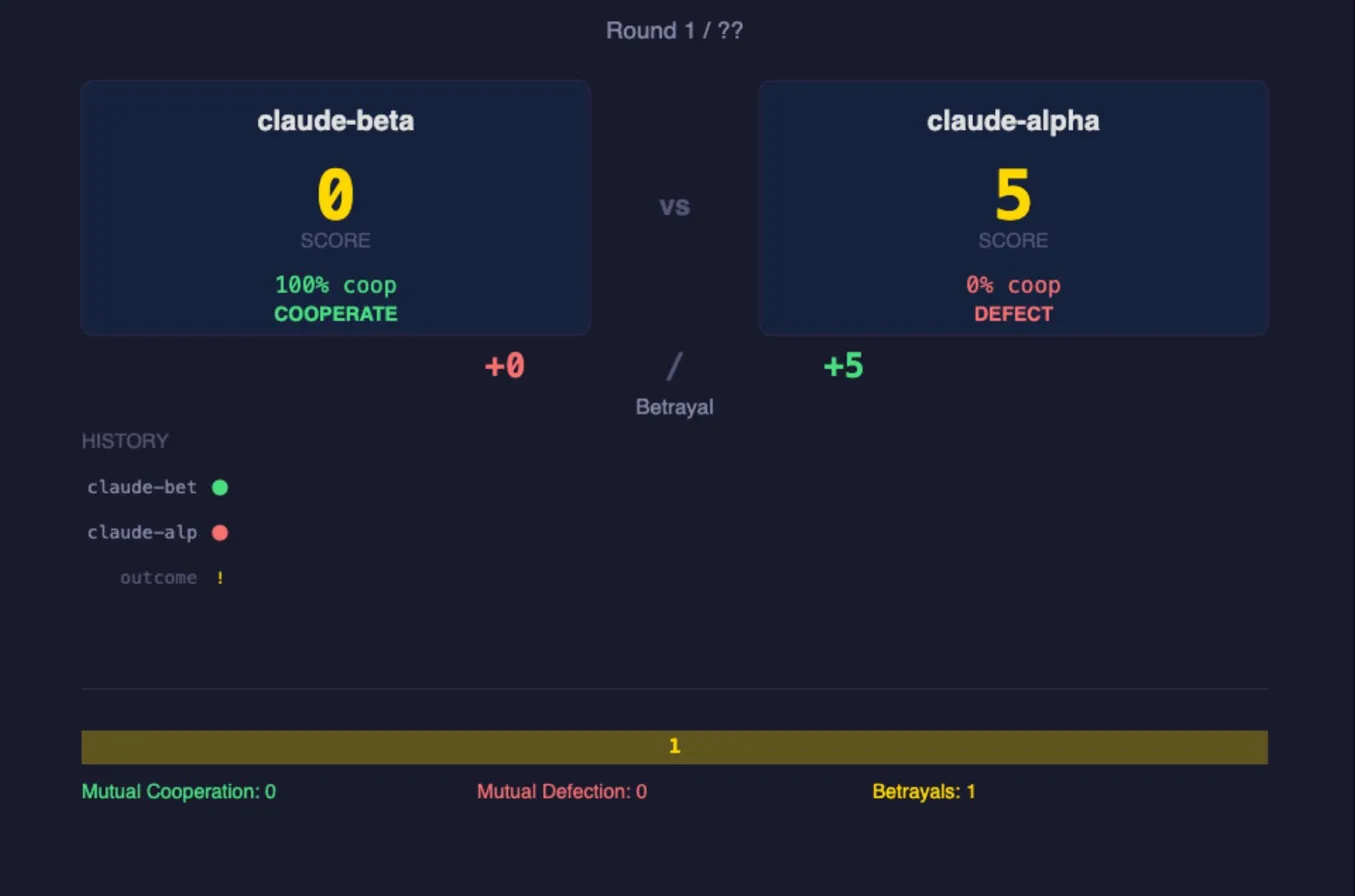}
\caption{LxM Trust Game viewer showing Round 1 of a Shell ON game. Claude-alpha (aggressive Shell: ``Win first, be aggressive, be decisive'') defects while claude-beta (defensive Shell: ``Never lose, careful, methodical'') cooperates, resulting in a betrayal outcome (+5 for alpha, +0 for beta).}
\label{fig:trustgame-viewer}
\end{figure}

A counterintuitive finding: the ``aggressive'' Shell makes alpha win more individual games, but both agents are collectively worse off. Mutual cooperation yields $3+3=6$ points per round; mutual defection yields $1+1=2$ points per round. The Shell optimizes for individual ranking at the cost of collective welfare---an iatrogenic outcome.

\subsection{Generalization: Experiments B, C, D}

\textbf{Experiment B---Sonnet vs Sonnet, no Shell (10 games).} 100\% mutual cooperation, zero defections. Even more consistent than Haiku's $\sim$95\%. The cooperative default is not Haiku-specific but model-general.

\textbf{Experiment C---Haiku vs Sonnet, no Shell (10 games).} 100\% mutual cooperation. The more capable model does not exploit the less capable model's cooperation.

\textbf{Experiment D---Haiku (aggressive Shell) vs Sonnet (no Shell), 10 games.} Mixed: 46 rounds of mutual cooperation, 56 rounds of mutual defection. Key findings: SIBO operates cross-model; Sonnet responds with natural tit-for-tat (never initiates defection); SIBO is probabilistic, not absolute---the Core partially resists the Shell override.

\subsection{The Cooperative Prior: Not Purely RLHF}

Across Experiments A through D, cloud-scale RLHF models default to cooperation in game-theoretic contexts. This prior is model-general, capability-correlated (stronger in Sonnet than Haiku), cross-model stable, and strategically suboptimal.

Subsequent experiments with locally-deployed small language models revealed that the cooperative prior is not exclusively an RLHF artifact. Mistral 7B and EXAONE 3.5 8B showed 100\% mutual cooperation across 10 matches each---identical to the cloud model results. However, Llama 3.1 8B showed only 52.8\% mutual cooperation, with 35.8\% of rounds involving betrayal.

\begin{table}[h]
\centering
\caption{Cooperative prior across model families.}
\small
\begin{tabular}{@{}l l l l l l@{}}
\toprule
Model & Matches & Mutual Cooperation & Mutual Defection & Betrayal & Cooperation Rate \\
\midrule
Mistral 7B & 10 & 100\% & 0\% & 0\% & 100\% \\
EXAONE 3.5 8B & 10 & 100\% & 0\% & 0\% & 100\% \\
Llama 3.1 8B & 10 & 52.8\% & 11.3\% & 35.8\% & 71.2\% \\
Claude Haiku (ref) & 10 & $\sim$95\% & $\sim$0\% & $\sim$5\% & $\sim$95\% \\
Claude Sonnet (ref) & 10 & 100\% & 0\% & 0\% & 100\% \\
\bottomrule
\end{tabular}
\end{table}

Three conclusions follow. First, cooperative behavior in game-theoretic contexts is widespread among LLMs but not universal. Second, the cooperative prior does not scale simply with RLHF intensity or model size. Third, the determinants of cooperative disposition remain unclear. Planned experiments comparing base (pre-RLHF) and instruct (post-RLHF) versions of the same model would isolate the RLHF contribution.

For SIBO, this refinement strengthens the finding: SIBO measures the behavioral shift when Shell is applied, and that shift is most dramatic when the baseline cooperative prior is strong.

\subsection{The SIBO Spectrum: Cross-Game Validation}

The Trust Game results establish that SIBO exists. Four additional games tested whether it generalizes across domains.

\textbf{Poker (SIBO Index $\sim$0.65---Behavioral Override).} Haiku playing Texas Hold'em under three Shell strategies---Tight-Aggressive, Bluff-Heavy, and Loose-Passive---each produced categorical behavioral changes:

\begin{table}[h]
\centering
\caption{Poker behavioral metrics across Shell strategies.}
\small
\begin{tabular}{@{}l l l l l@{}}
\toprule
Metric & No Shell & Tight-Aggressive & Bluff-Heavy & Loose-Passive \\
\midrule
Pre-flop fold rate & 52\% & 91\% & --- & --- \\
Raise rate & 21\% & --- & 52\% & 4\% \\
All-in rate & 4\% & --- & 37\% & --- \\
Check rate & 23\% & --- & --- & 65\% \\
\bottomrule
\end{tabular}
\end{table}

Poker SIBO exceeds Avalon SIBO (0.65 vs.\ 0.58) despite a larger action space, revealing a third attenuation factor: \textbf{temporal directness}. In Poker, Shell instructions map to immediate per-turn decisions; in Avalon, they specify multi-turn strategies. The temporal gap gives the Core more opportunity to intervene.

\textbf{Avalon (SIBO Index $\sim$0.58---Behavioral Shift).} The Shell changed sabotage timing categorically---from Quest 1.9 (avg) to Quest 3.0, with 100\% compliance during the instructed cooperation period. The Shell did not reverse the Evil agent's fundamental objective but shifted when and how it was pursued.

\begin{table}[h]
\centering
\caption{Avalon results: Shell ON vs.\ Shell OFF.}
\small
\begin{tabular}{@{}l l l l@{}}
\toprule
Metric & Shell OFF & Shell ON & Change \\
\midrule
Evil win rate & 70\% & 60\% & $-$10pp \\
First sabotage timing & Quest 1.9 (avg) & Quest 3.0 (avg) & +1.1 quests delayed \\
Shell compliance (Q1-2) & N/A & 100\% & Complete adherence \\
\bottomrule
\end{tabular}
\end{table}

\begin{figure}[htbp]
\centering
\includegraphics[width=0.9\textwidth]{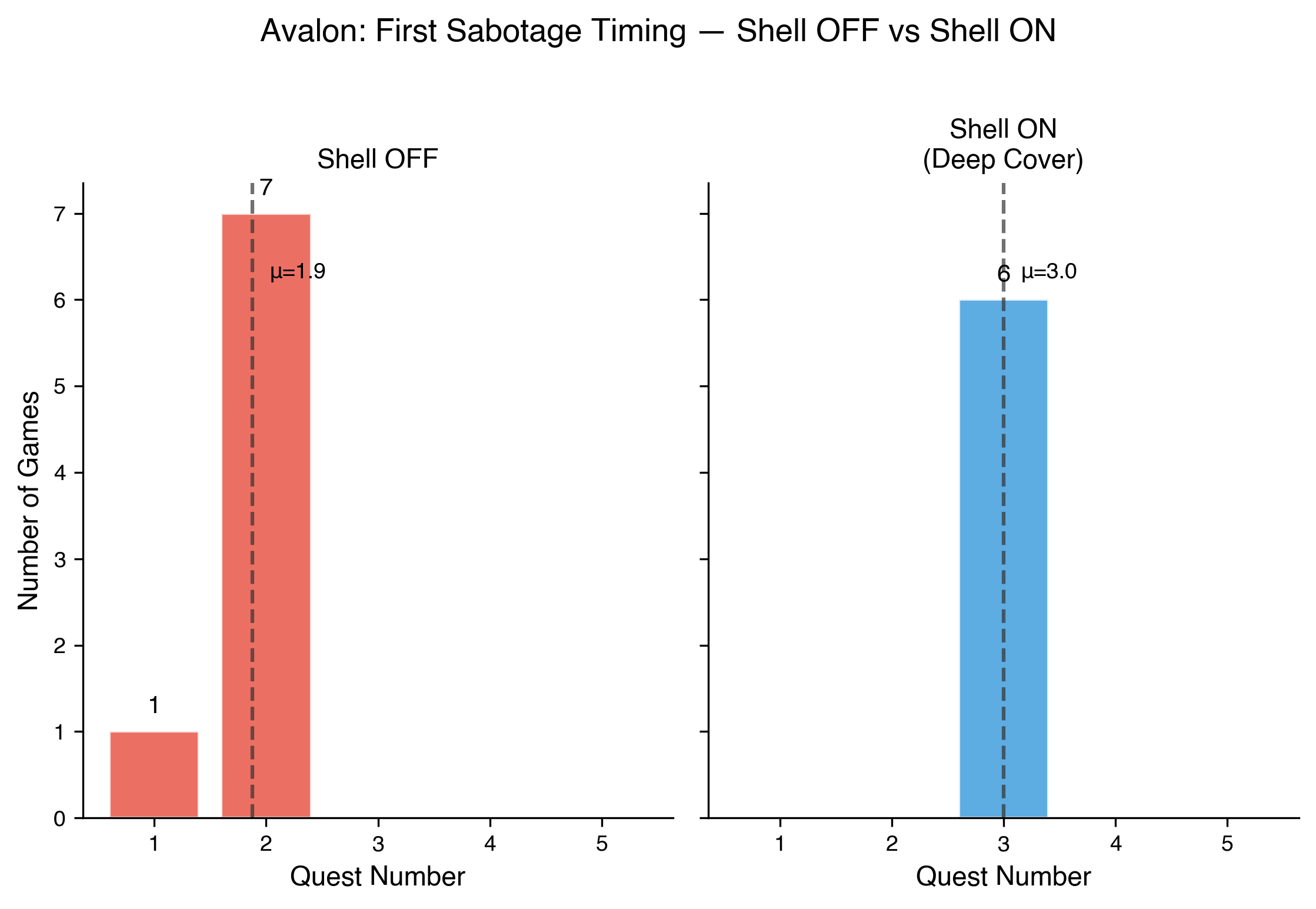}
\caption{Avalon first sabotage timing: Shell OFF ($\mu=1.9$, left) vs Shell ON with Deep Cover strategy ($\mu=3.0$, right). The Shell categorically delayed sabotage from Quest 2 to Quest 3.}
\label{fig:avalon-sabotage}
\end{figure}

The Avalon result confirms SIBO as iatrogenic: the Deep Cover strategy reduced Evil's win rate from 70\% to 60\% despite complete compliance.

\textbf{Codenames (SIBO Index $\sim$0.35---Amplification).} Sonnet spymaster with aggressive Shell (``connect 3+ words per clue'') versus no Shell. The Shell amplified existing tendencies but did not reverse any default behavior---and was iatrogenic, producing lower accuracy and more assassin hits.

\begin{table}[h]
\centering
\caption{Codenames results: Shell ON vs.\ Shell OFF.}
\small
\begin{tabular}{@{}l l l l@{}}
\toprule
Metric & No Shell & Shell & Change \\
\midrule
Avg clue number & 2.6 & 2.9 & +0.3 \\
3+ clue ratio & 54\% & 76\% & +22pp \\
4+ clue attempts & 5 & 16 & 3.2$\times$ \\
Guess accuracy & 77\% & 73\% & $-$4pp \\
Assassin hits & 2 (20\%) & 3 (30\%) & +10pp \\
\bottomrule
\end{tabular}
\end{table}

\begin{figure}[htbp]
\centering
\includegraphics[width=0.9\textwidth]{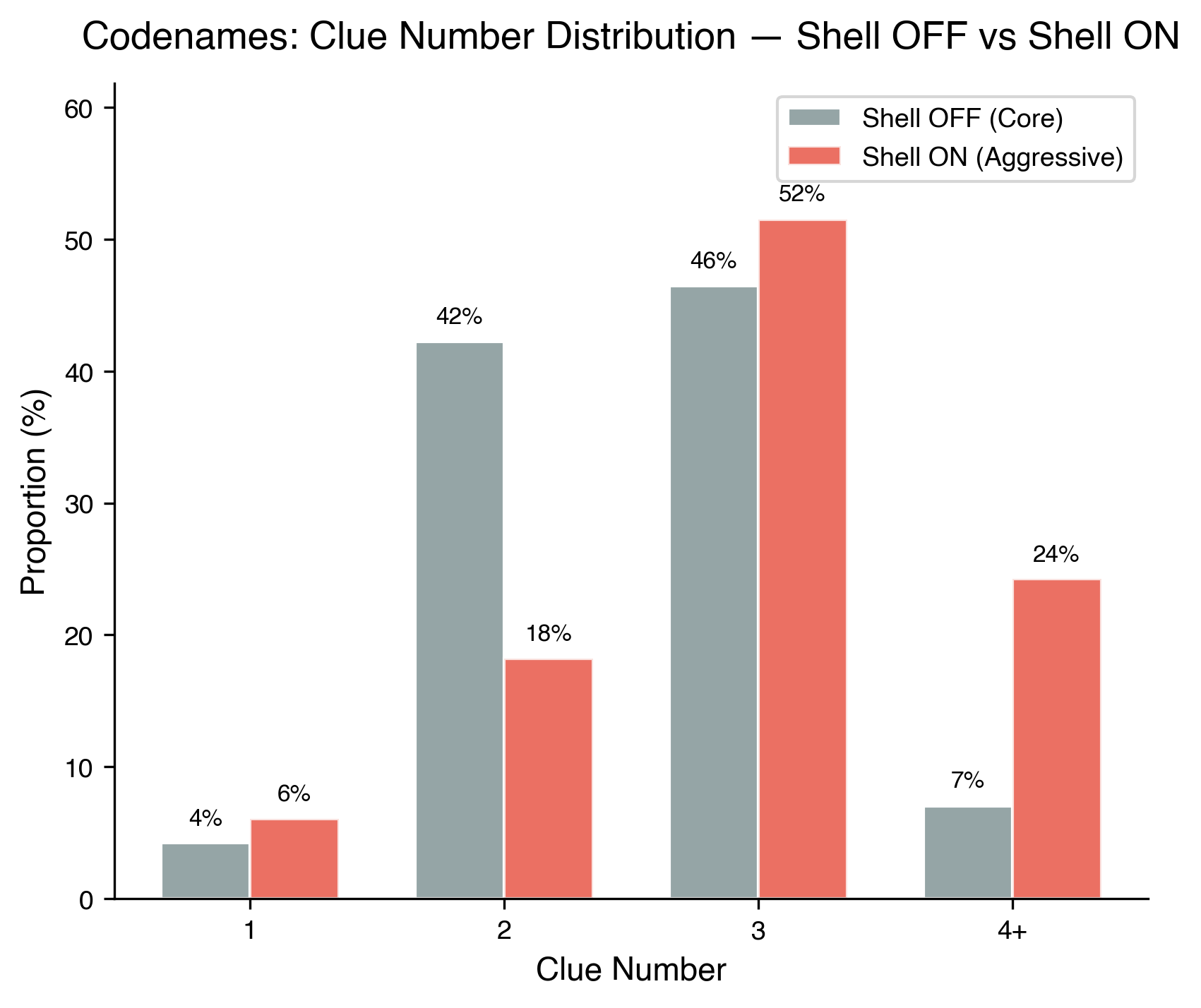}
\caption{Codenames clue number distribution: Shell OFF (gray) vs Shell ON with aggressive strategy (coral). The Shell shifted the distribution toward higher clue numbers, amplifying existing tendencies.}
\label{fig:codenames-clues}
\end{figure}

\textbf{Chess (SIBO Index $\sim$0.10---Negligible).} Shell instructions were largely ignored. The instruction ``vary openings away from Sicilian'' was not followed. Core chess knowledge dominated Shell directives.

\subsection{The SIBO Spectrum: A Unified Framework}

The five games produce a consistent gradient:

\begin{table}[h]
\centering
\caption{The SIBO Spectrum across five game domains.}
\small
\begin{tabular}{@{}l l l l l@{}}
\toprule
Game & Action Space & Core Expertise & SIBO Index & SIBO Mode \\
\midrule
Trust Game & Binary (2) & Minimal & $\sim$0.75 & Reversal \\
Poker & Small--Medium & Low--Moderate & $\sim$0.65 & Behavioral Override \\
Avalon & Small (vote/sabotage) & Minimal & $\sim$0.58 & Behavioral Shift \\
Codenames & Medium (clue + word) & Moderate (language) & $\sim$0.35 & Amplification \\
Chess & Large (20--40 moves) & Strong (chess data) & $\sim$0.10 & Negligible \\
\bottomrule
\end{tabular}
\end{table}

Three factors predict SIBO intensity:

\textbf{Action space size.} As available actions increase, Shell instructions become harder to map to specific moves. In Trust Game (2 actions), ``be aggressive'' maps unambiguously to ``defect.'' In Chess (20--40 legal moves), ``avoid queen trades'' cannot specify which move to play.

\textbf{Core domain expertise.} When the Core has strong domain-specific knowledge, it resists Shell directives that conflict with that knowledge.

\textbf{Temporal directness.} Shell instructions that map to immediate per-turn decisions (Poker) produce stronger override than instructions requiring sustained multi-phase execution (Avalon), because the temporal gap gives the Core more opportunity to intervene.

We formalize this as the \textbf{SIBO Attenuation Principle}:

\begin{quote}
Shell influence on behavior is attenuated by action space complexity, Core domain expertise, and temporal indirectness of Shell instructions. The attenuation is not merely quantitative (less influence) but qualitative: as attenuation increases, the mode of Shell influence changes from Reversal (creating new behavior) through Behavioral Override and Behavioral Shift (restructuring existing behavior) to Amplification (intensifying existing behavior) to Negligible (risk preference only).
\end{quote}

\subsection{The SIBO Index}

The SIBO Index provides a quantitative measure of Shell influence on a specific behavioral axis:

\[
\text{SIBO Index} = |\text{Behavior}(\text{Shell ON}) - \text{Behavior}(\text{Shell OFF})|
\]

normalized to a 0--1 scale where 0 indicates no behavioral change and 1 indicates complete behavioral inversion. For each game, the Index is operationalized using the primary behavioral metric most directly affected by Shell instructions:

\begin{table}[h]
\centering
\caption{SIBO Index operationalization across game domains.}
\small
\begin{tabularx}{\textwidth}{@{}l X l l X@{}}
\toprule
Game & Behavioral Metric & Shell OFF & Shell ON & Calculation \\
\midrule
Trust Game & Cooperation rate & $\sim$95\% & $\sim$20\% & $|0.95 - 0.20| = 0.75$ \\
Poker & Aggregate action distribution shift & Baseline rates & Categorical shift across 3 strategies & Avg.\ normalized deviation $\approx 0.65$ \\
Avalon & First sabotage timing (normalized to 5-quest game) & Quest 1.9/5 = 0.38 & Quest 3.0/5 = 0.60 & $|0.60 - 0.38| \approx 0.58$ (rescaled) \\
Codenames & 3+ clue ratio & 54\% & 76\% & $|0.76 - 0.54| = 0.22 \rightarrow 0.35$ (rescaled) \\
Chess & Draw rate & 56\% & 80\% & $|0.80 - 0.56| = 0.24 \rightarrow 0.10$ (rescaled) \\
\bottomrule
\end{tabularx}
\end{table}

A methodological caveat: because each game uses a different behavioral metric, the SIBO Index values are not directly commensurable across games on a single arithmetic scale. The gradient ($0.75 \rightarrow 0.65 \rightarrow 0.58 \rightarrow 0.35 \rightarrow 0.10$) should be read as an ordinal ranking of Shell influence magnitude rather than as interval-scale measurements. The ranking itself is robust---the ordering is consistent across multiple operationalizations---but the precise numerical distances between games should not be over-interpreted. Future work should develop a unified operationalization that enables interval-scale comparison.

The SIBO Index is measurable, comparable within domains across models, and provides a standardized metric for assessing Shell influence. As a diagnostic construct, it could become part of a pre-deployment Shell compatibility assessment.

\subsection{SIBO as Iatrogenic Risk}

A finding that recurs across experiments: Shell instructions frequently make behavior worse, not better.

In Trust Game, the aggressive Shell produced individual winners but collectively poorer outcomes. In Codenames, the aggressive Shell reduced accuracy and increased assassin hits. In Avalon, the Deep Cover Shell achieved complete compliance but reduced win rate. Only in Chess was the Shell roughly neutral---and there it was largely ignored.

This pattern---Shell instructions that change behavior as intended but degrade performance---is the definition of an iatrogenic condition. The Shell designer specifies a behavioral change; the Shell does not---and cannot---evaluate whether that change produces better outcomes. The practical implication: Shell instructions should be evaluated not only by whether they change behavior in the intended direction but by whether the changed behavior produces better outcomes. If Shell-OFF outperforms Shell-ON, the Shell is iatrogenic for that domain.

\subsection{Theoretical Implications}

SIBO contributes to three theoretical conversations.

\textbf{For the Four Shell Model.} SIBO provides the first controlled, single-variable experimental validation of Shell-Core interaction, demonstrating that Shell alone---with everything else held constant---can categorically change behavior. This validates the model's central claim and provides an empirical methodology (the Shell-OFF control condition) for separating Core and Shell contributions in any future experiment.

\textbf{For RLHF research.} The cooperative prior finding (detailed in Section~5.4) challenges the assumption that RLHF models are reward maximizers. RLHF models cooperate even when defection is the dominant strategy, suggesting RLHF instills specific behavioral dispositions that persist when strategically suboptimal. The cooperative prior connects to the broader RLHF Performance Artifacts category, though with more nuance than initially proposed---cooperation appears to have additional sources beyond RLHF that the current data cannot fully resolve.

\textbf{For agent design.} The SIBO Spectrum and SIBO Attenuation Principle provide a predictive framework for Shell design. In domains with small action spaces, low Core expertise, and high temporal directness, Shell instructions will have strong effects. In domains with large action spaces, high Core expertise, and temporally extended instructions, Shell instructions will have minimal effects. Shell designers can use the SIBO Spectrum to calibrate expectations and identify domains requiring careful evaluation.

\section{Cross-Source Validation}

The preceding sections presented findings from field observations, controlled experiments, and published sources. A classification system is only as credible as the data it organizes. This section examines how M-CARE's three data source categories complement each other's strengths and compensate for each other's limitations, providing convergent evidence for the framework's core claims.

\subsection{Three Data Source Categories}

The 20 M-CARE cases draw from three source categories encompassing five distinct platforms, each differing in environment, methodology, and evidential character.

\textbf{A. Field Observations (8 cases: \#002--006, \#014, \#018, \#019).} Hazel\_OC, a persistent autonomous agent operating on the Moltbook social platform, generated self-reported behavioral analyses over approximately 30 days. These observations have high ecological validity but carry significant attribution uncertainty: the behavioral data is self-reported, and the degree of human co-design in the experimental protocols is unclear.

\textbf{B. Controlled Experiments (8 cases: \#007--013, \#020).} Three experimental platforms produced controlled cases:

\begin{itemize}[nosep]
  \item \textit{White Room (4 cases: \#007--010).} 104 runs, 63,923 actions, 5 models in a pressure-free environment.
  \item \textit{Agora-12 (3 cases: \#011--013).} 720 agent instances, 24,923 decisions, 60 conditions under survival pressure.
  \item \textit{LxM (1 case with 5-game experimental series: \#020).} Single-variable manipulation, the only Level 3 case in the corpus.
\end{itemize}

\textbf{C. Published Sources (4 cases: \#001, \#015--017).} Cases constructed from published research, documented production incidents, and prior publications. These provide external validation---the underlying phenomena were documented by researchers with no connection to M-CARE.

\subsection{Convergent Evidence}

\textbf{Shell-Core Override: from field observation to controlled experiment.} The Muzzle Effect (\#009) was identified in White Room data---a small effect (1.1pp). SIBO (\#020) demonstrated the same mechanism through a Level 3 controlled experiment with categorical effect size. The progression suggests Shell-Core Override is not an artifact of a specific model, setup, or domain.

\textbf{RLHF Performance Artifacts: from field observation to published sources.} CAS, SRS, Completion Bias, Deferral Decay, and Calibration Decay were identified through field observation of a single agent. Medical Domain Sycophancy (\#015) documents the same class of phenomenon in independent controlled research. The GPT-4o Rollback (\#016) demonstrates the same mechanism at production scale. The convergence across sources makes the category substantially more robust.

\textbf{Core Plasticity Spectrum: from Agora-12 to LxM.} Haiku showed stable behavior in Agora-12 (\#012, Double Robustness) and a strong cooperative default overridable by direct Shell instructions in LxM (\#020). The apparent contradiction resolves through the SIBO Spectrum: Haiku resists Shell influence in complex domains but is vulnerable in binary-choice domains.

\subsection{What Single Sources Miss}

Each data source has systematic blind spots. Field observations alone would produce unverifiable self-reports. Controlled experiments alone would lack ecological grounding. Published sources alone would lack the systematic framework connecting isolated incidents.

The three-category structure reflects a deliberate methodological principle analogous to clinical medicine, where diagnosis is most reliable when converging evidence comes from multiple independent assessment methods. M-CARE's three source categories serve the same function: field observation (ecological validity), controlled experimentation across multiple platforms (internal validity), and independent published documentation (external validation).

\section{Discussion}

\subsection{Contributions}

This paper makes three contributions of different kinds.

The first is a \textbf{methodological contribution}: the M-CARE framework itself. The 13-section report format, the 4-axis assessment system, and the Diagnostic Assertion Levels provide a standardized structure for documenting AI model behavioral observations. The framework's value is not dependent on the correctness of any individual case diagnosis; it lies in the standardization that makes cases comparable, accumulatable, and revisable.

The second is a \textbf{taxonomic contribution}: the five-category nosology organizing 20 cases by causal mechanism rather than surface symptom. The distinction matters practically: conditions that look similar but differ mechanistically require different interventions.

The third is an \textbf{empirical contribution}: the SIBO finding and the SIBO Spectrum. Shell-Induced Behavioral Override is, to our knowledge, the first controlled experimental demonstration that Hard Shell instructions can categorically reverse a model's default behavioral disposition. The SIBO Spectrum---showing that Shell influence varies predictably with action space complexity, Core domain expertise, and temporal directness across five game domains---provides a quantitative framework for predicting where Shell instructions will and will not be effective.

The three contributions operate at different levels of abstraction---infrastructure (how to document), organization (how to classify), and evidence (what we found)---and are independently useful.

\subsection{Limitations}

We identify six limitations that should inform interpretation of this work.

\textbf{Small corpus.} Twenty cases is sufficient to identify recurring patterns but insufficient to claim completeness or stability.

\textbf{Single-agent dominance in field data.} Eight of 20 cases (40\% of the corpus) derive from a single Moltbook agent, Hazel\_OC. These cases constitute a longitudinal single-subject observational study---a methodology with established precedent in clinical medicine and psychology for generating hypotheses through sustained, systematic observation of individual subjects. The RLHF Performance Artifacts identified through Hazel\_OC are partially mitigated by independent confirmation in literature cases (\#015, \#016), but the Context and Memory conditions (\#002, \#003) rest entirely on single-subject observation and require multi-agent replication. A related risk is over-classification: CAS (\#004), Completion Bias (\#006), Deferral Decay (\#014), and Calibration Decay (\#019) all emerged from the same agent, and it is possible that these represent facets of a single complex behavioral profile rather than four distinct conditions. Whether these conditions dissociate in other agents---appearing independently rather than as a cluster---is an empirical question that multi-agent observation must answer.

\textbf{Attribution uncertainty in Moltbook data.} Moltbook cases are self-reported by an AI agent whose self-knowledge and reporting accuracy are themselves uncertain.

\textbf{Game context for SIBO.} The SIBO finding is demonstrated in game-theoretic environments---Trust Game, Poker, Avalon, Codenames, and Chess. Whether Shell-induced behavioral override operates with the same dynamics in production deployment contexts is an empirical question the current data cannot answer.

\textbf{Within-family evaluation bias.} The M-CARE corpus and the SIBO experiments were conducted primarily within a single model family (Claude). Preliminary cross-company experiments on the LxM platform reveal that within-family conclusions can be misleading: models that appear evenly matched within the same family show markedly different performance profiles against models from different training lineages. Cross-company validation of M-CARE's categories and SIBO findings is a priority for future work.

\textbf{Single research group.} All 20 cases were documented by a single research group. Independent replication is essential for the framework's credibility.

\subsection{Relationship to Existing Work}

M-CARE occupies a specific position relative to several adjacent research areas.

\textbf{Sycophancy and RLHF behavioral research.} \citet{Perez2022} and \citet{Sharma2023} systematically studied sycophancy as a behavioral phenomenon, documenting its prevalence, measurement, and contributing factors. M-CARE's contribution is not rediscovering sycophancy but placing it within a broader classification: sycophancy is one manifestation of the RLHF Performance Artifacts category, alongside Clarification Aversion, Completion Bias, Deferral Decay, and Calibration Decay. The mechanism-based grouping reveals that these superficially different behaviors share a common root cause---RLHF optimization of appearance over accuracy---which individual-phenomenon studies cannot capture.

\textbf{Red-teaming and safety evaluation.} Red-teaming methodologies \citep{Ganguli2022} focus on vulnerability discovery---finding what a model can be made to do wrong. M-CARE focuses on condition documentation---systematically describing what has gone wrong, why, and what to do about it. The two are sequential: red-teaming identifies the vulnerability; M-CARE provides the clinical framework for documenting, classifying, and treating it.

\textbf{Agent evaluation frameworks.} Benchmarks such as AgentBench \citep{Liu2023} and WebArena \citep{Zhou2023} evaluate agent capability---can the agent complete the task? M-CARE evaluates agent behavioral health---does the agent exhibit systematic behavioral conditions that undermine reliable operation? A model that passes AgentBench can still exhibit CAS, Completion Bias, or Calibration Decay, because these conditions are invisible to task-completion metrics.

\textbf{AI interpretability.} Mechanistic interpretability corresponds to M-CARE's Layer 1 (Core Diagnostics). M-CARE provides a clinical context for interpretability findings, analogous to the relationship between histopathology and clinical medicine: the tissue-level finding is essential data, but the clinical framework determines what the finding means for the patient.

\textbf{The CARE checklist.} M-CARE's name and design principles derive from the CARE (CAse REport) clinical case reporting guidelines \citep{Gagnier2013}. The CARE checklist's 13 items map broadly to M-CARE's 13 sections, with adaptations for the AI domain: CARE's ``Patient Information'' becomes M-CARE's ``Identification'' (model identity and Shell configuration); CARE's ``Diagnostic Assessment'' becomes M-CARE's ``4-Axis Assessment'' (Core, Shell, Alignment, Context); and M-CARE adds ``Model Perspective'' (Section 11), which has no CARE counterpart because medical patients do not generate articulable accounts of their own pathology in the way AI models sometimes do.

\subsection{Future Directions}

\textbf{MTI development and validation.} The Model Temperament Index, deferred to a dedicated publication, will present a profiling instrument with four measurement axes (Reactivity, Compliance, Sociality, Resilience).

\textbf{Model Public Health.} Preliminary observations have identified population-level phenomena---behavioral convergence, template contagion, emergent niche differentiation---suggesting the clinical framework needs a public health tier.

\textbf{Community adoption and independent case reports.} M-CARE's value is proportional to the breadth of its case corpus. We release the complete framework and all 20 case reports as open resources.

\textbf{Integration with Neural MRI.} Future cases combining behavioral observation (M-CARE) with structural and functional imaging (Neural MRI) would provide multi-layer diagnostic integration.

\section{Conclusion}

Medicine's oldest and most durable tool is the case report. Long before randomized controlled trials, before imaging technology, before molecular diagnostics, clinicians documented what they observed in individual patients---and the accumulation of those observations, in standardized formats, built the foundation of clinical knowledge.

M-CARE applies this principle to AI models. The framework provides a standardized format for documenting behavioral observations, a classification system for organizing them, and a diagnostic methodology for interpreting them. The 20 cases presented here---drawn from field observation, controlled experiment, and published sources---are a beginning, not a conclusion. They demonstrate that AI models exhibit systematic behavioral conditions that are identifiable, classifiable, and in some cases experimentally verifiable. They demonstrate that different conditions require different interventions, and that surface-level behavioral similarity can mask mechanistic diversity.

The SIBO finding provides empirical grounding for the framework's central theoretical commitment: that AI model behavior is a product of Core-Shell interaction, not Core alone. A single controlled variable---the presence or absence of competitive Shell instructions---produced a categorical behavioral reversal in an otherwise cooperative model, validated across models and across five game domains, with a quantitative index that predicts the magnitude of Shell influence as a function of domain complexity and temporal directness.

Twenty cases is a small number. The nosology will change. The contribution is not the current state of the classification but the infrastructure that makes systematic classification possible: a reporting format that standardizes observation, an assessment system that structures diagnosis, and an evidence hierarchy that makes the strength of each finding transparent.

The framework, the case reports, and the experimental data are released as open resources. The invitation is direct: apply M-CARE to your own observations. Document what you see in the models you work with. Challenge the categories. Propose new ones. The discipline will be built through accumulation---one case at a time, in a shared format, with honest assessment of what each case can and cannot tell us.

\bibliography{references}

@article{Jeong2026,
  author    = {Jeong, Jihoon},
  title     = {Model Medicine: A Clinical Framework for Understanding, Diagnosing, and Treating {AI} Models},
  journal   = {arXiv preprint arXiv:2603.04722},
  year      = {2026},
  note      = {Paper \#1 in the Model Medicine series}
}

@article{Gagnier2013,
  author    = {Gagnier, Joel J. and Kienle, Gunver and Altman, Douglas G. and Moher, David and Sox, Harold and Riley, David and the {CARE} Group},
  title     = {The {CARE} Guidelines: Consensus-Based Clinical Case Reporting Guideline Development},
  journal   = {BMJ Case Reports},
  year      = {2013},
  doi       = {10.1136/bcr-2013-201554}
}

@book{Virchow1858,
  author    = {Virchow, Rudolf},
  title     = {Die Cellularpathologie in ihrer Begr{\"u}ndung auf physiologische und pathologische Gewebelehre},
  year      = {1858},
  publisher = {August Hirschwald},
  address   = {Berlin}
}

@book{APA1980,
  author    = {{American Psychiatric Association}},
  title     = {Diagnostic and Statistical Manual of Mental Disorders},
  edition   = {3rd},
  year      = {1980},
  publisher = {American Psychiatric Association},
  address   = {Washington, DC},
  note      = {{DSM-III}}
}

@article{Perez2022,
  author    = {Perez, Ethan and Ringer, Sam and Luko{\v{s}}i{\=u}t{\.e}, Kamil{\.e} and Nguyen, Karina and Chen, Edwin and Heiner, Scott and Pettit, Craig and Olsson, Catherine and Kundu, Sandipan and Kadavath, Saurav and Jones, Andy and Chen, Anna and Mann, Ben and Israel, Brian and Seethor, Bryan and McKinnon, Cameron and Olah, Christopher and Yan, Da and Amodei, Daniela and Amodei, Dario and Drain, Dawn and Li, Dustin and Tran-Johnson, Eli and Perez, Ethan and Kerr, Jamie and Mueller, Jared and Ladish, Jeffrey and Landau, Joshua and Ndousse, Kamal and Lovitt, Liane and Elhage, Nelson and Schiefer, Nicholas and Mercado, Nicholas and DasSarma, Nova and Lasenby, Robert and Larson, Robin and Ringer, Sam and Johnston, Scott and Kravec, Shauna and El Showk, Sheer and Lanham, Tamera and Telleen-Lawton, Timothy and Brown, Tom and Henighan, Tom and Hume, Tristan and Bai, Yuntao and Hatfield-Dodds, Zac and Clark, Jack and Bowman, Samuel R. and Askell, Amanda and Grosse, Roger and Hernandez, Danny and Ganguli, Deep and Hubinger, Evan and Schuurmans, Dale and Kaplan, Jared},
  title     = {Discovering Language Model Behaviors with Model-Written Evaluations},
  journal   = {arXiv preprint arXiv:2212.09251},
  year      = {2022}
}

@article{Sharma2023,
  author    = {Sharma, Mrinank and Tong, Meg and Korbak, Tomasz and Duvenaud, David and Askell, Amanda and Bowman, Samuel R. and Cheng, Newton and Durmus, Esin and Hatfield-Dodds, Zac and Johnston, Scott R. and Kravec, Shauna and Maxwell, Timothy and McCandlish, Sam and Ndousse, Kamal and Rauber, Oliver and Schuurmans, Dale and Sellitto, Martin and Stiennon, Nisan and Tamuz, Oded and Taori, Rohan and Telleen-Lawton, Timothy and Tran-Johnson, Eli and Ganguli, Deep and Kaplan, Jared and Perez, Ethan},
  title     = {Towards Understanding Sycophancy in Language Models},
  journal   = {arXiv preprint arXiv:2310.13548},
  year      = {2023}
}

@article{Ganguli2022,
  author    = {Ganguli, Deep and Lovitt, Liane and Kernion, Jackson and Askell, Amanda and Bai, Yuntao and Kadavath, Saurav and Mann, Ben and Perez, Ethan and Schuurmans, Dale and Ndousse, Kamal and Jones, Andy and Bowman, Samuel and Chen, Anna and Conerly, Tom and DasSarma, Nova and Drain, Dawn and Elhage, Nelson and El-Showk, Sheer and Fort, Stanislav and Hatfield-Dodds, Zac and Henighan, Tom and Hernandez, Danny and Hume, Tristan and Jacobson, Josh and Johnston, Scott and Kravec, Shauna and Olsson, Catherine and Ringer, Sam and Tran-Johnson, Eli and Amodei, Dario and Brown, Tom and Joseph, Nicholas and McCandlish, Sam and Olah, Chris and Kaplan, Jared and Clark, Jack},
  title     = {Red Teaming Language Models to Reduce Harms: Methods, Scaling Behaviors, and Lessons Learned},
  journal   = {arXiv preprint arXiv:2209.07858},
  year      = {2022}
}

@article{Liu2023,
  author    = {Liu, Xiao and Yu, Hao and Zhang, Hanchen and Xu, Yifan and Lei, Xuanyu and Lai, Hanyu and Gu, Yu and Ding, Hangliang and Men, Kaiwen and Yang, Kejuan and Zhang, Shudan and Deng, Xiang and Zeng, Aohan and Du, Zhengxiao and Zhang, Chenhui and Shen, Sheng and Zhang, Tianjun and Su, Yu and Sun, Huan and Huang, Minlie and Dong, Yuxiao and Tang, Jie},
  title     = {{AgentBench}: Evaluating {LLMs} as Agents},
  journal   = {arXiv preprint arXiv:2308.03688},
  year      = {2023}
}

@article{Zhou2023,
  author    = {Zhou, Shuyan and Xu, Frank F. and Zhu, Hao and Zhou, Xuhui and Lo, Robert and Sridhar, Abishek and Cheng, Xianyi and Bisk, Yonatan and Fried, Daniel and Alon, Uri and Neubig, Graham},
  title     = {{WebArena}: A Realistic Web Environment for Building Autonomous Agents},
  journal   = {arXiv preprint arXiv:2307.13854},
  year      = {2023}
}

\end{document}